\documentclass[preprint,showpacs,preprintnumbers,amsmath,amssymb,showkeys,chaos]{revtex4-1}

\usepackage{graphicx}
\usepackage{dcolumn}
\usepackage{bm}
\usepackage{epsfig}
\usepackage{amsmath}

\begin{document}
\title{Scroll wave drift along steps, troughs and corners}
\author{Hua Ke}
\author{Zhihui Zhang}
\author{Oliver Steinbock}
\affiliation{Florida State University, Department of Chemistry and Biochemistry, Tallahassee, FL 32306-4390}



\begin{abstract}
Three-dimensional excitable systems can create nonlinear scroll waves that rotate around one-dimensional phase singularities. Recent theoretical work predicts that these filaments drift along step-like height variations. Here we test this prediction using experiments with thin layers of the Belousov-Zhabotinsky reaction. We observe that over short distances scroll waves are attracted towards the step and then rapidly commence a steady drift along the step line. The translating filaments always reside in the shallow subsystem and terminate on the step plateau near the edge. Accordingly filaments in the deep subsystem initially collide with and shorten at the step wall. The drift speeds obey the predicted proportional dependence on the logarithm of the height ratio and the direction depends on the vortex chirality. We also observe drift along the perimeter of rectangular plateaus and find that the filaments perform sharp turns at the corners. In addition, we investigate rectangular troughs for which vortices of equal chirality can drift in different directions. The latter two effects are reproduced in numerical simulations with the Barkley model. The simulations show that narrow troughs instigate scroll wave encounters that induce repulsive interaction and symmetry breaking. Similar phenomena could exist in the geometrical complicated ventricles of the human heart where reentrant vortex waves cause tachycardia and fibrillation.
\end{abstract}
\maketitle

\section{Lead Paragraph}

Bistability, excitability, oscillations, and chaos are common features in systems far from the thermodynamic equilibrium. In spatially extended systems, these dynamics give rise to symmetry-breaking selforganization. Well-known examples include Turing patterns, traveling waves of excitation, and spatio-temporal chaos. Nearly all of these phenomena are found in living systems where they generate the order (or disorder) needed to bridge the length-scale divide between the molecular and the macroscopic world. To understand these processes, it is important to consider the impact of system geometry, confinement, and heterogeneity because all of these factors are intrinsic to living systems. For the example of a chemical reaction-diffusion medium and a related activator-inhibitor model, we show that already simple step-like variations in system height can cause excitation vortices to move along the contour of the step. These findings also confirm recent theoretical predictions obtained from an asymptotic theory and the vortices' characteristic response functions.

\section{Introduction}

Rotating spiral waves are a common feature in a diverse spectrum of dissipative systems and have attracted considerable research interest \cite{Mikhailov06, Vanag01, book98}. Examples include chemical reaction-diffusion media such as the homogeneous Belousov-Zhabotinsky (BZ) reaction \cite{winfree72}, BZ-based water-in-oil microemulsions \cite{epstein03}, and the chlorine dioxide-iodine malonic acid (CDIMA) reaction \cite{epstein08}. Furthermore, there are numerous biological systems that create these excitation waves to communicate information over macroscopic distances. This signal relay has been reported for single cells, cell populations, organs, and insect colonies. For instance, the slime mold \textit{Dictyostelium discoideum}, a widely studied model organism in developmental biology, generates traveling waves of cyclic-AMP concentration to selforganize its aggregation \cite{dicty} and giant honey bees have developed a communal response mechanism against predators that can result in spiral waves \cite{bees}. Also in the human body, spiral waves are involved in important processes ranging from cardiac arrhythmia to uterine contractions in child birth \cite{heart1,uterus}.

A two-dimensional excitation vortex is organized around a rotating reaction front that, due to its small width, can be described as a continuous curve extending from the spiral tip in center of the pattern outwards. In many cases, this curve traces an Archimedean spiral of constant pitch \cite{ost92}. Furthermore, the spiral tip is a phase singularity in which iso-concentration lines of the key reactants (e.g. an activator and an inhibitor species) intersect. This special feature of the spiral core also implies that spirals can only be created and destroyed as pairs of opposite chirality \cite{schutze92}. The location of the spiral core is constant in most unperturbed reaction-diffusion systems; however, vortex drift can arise from spatial gradients, temporal forcing, nearby heterogeneities such as system boundaries, and the short-range interaction between spiral tips. Examples for these cases have been investigated in numerous experimental systems and rely for BZ experiments on external perturbations such as electric fields, temperature gradients, and tailored light signals \cite{schutze92,zykov94}. Most of these studies reveal a particle-like behavior of the spiral core in which the pattern chirality constitutes a topological charge. For instance, spiral waves exposed to constant, parallel electric fields drift in field direction with perpendicular velocity components that depend on the sense of rotation \cite{schutze92}.

Our study focuses on related drift phenomena in three-dimensional excitable systems where wave rotation is organized by one-dimensional phase singularities. These curves and their surrounding wave fields are called filaments and scroll waves, respectively \cite{keenertyson,dutta10,hauser13}. Scroll waves can be considered a continuum of spiral waves for which the spiral tips extend along the filament. Changes in the rotation phase along this backbone manifest themselves as twist. The filament itself is not necessarily static but moves according to its local curvature and other dynamic parameters (e.g. strong twist). In the simplest cases, the speed of this motion is proportional to the local filament curvature and occurs in normal direction to the filament. The corresponding proportionality constant is known as the filament tension $\alpha$ \cite{biktashev94}. This system-specific constant can be positive or negative and induces curve-shrinking dynamics or turbulent filament growth, respectively. Motion in binormal direction occurs only in systems where the diffusion coefficients of the activator and control species are different. Accordingly, this motion is absent in the complex Ginzburg-Landau equation and very slow in BZ systems \cite{cgle}. Recent theoretical and experimental studies also documented a filament rigidity which is a fourth-order term in the underlying kinematic equations of filament motion \cite{nakouzi14}.

The externally induced change of scroll wave filaments is, by comparison to the perturbation of spiral waves, a widely understudied topic. Nonetheless, Vinson et al. reported that temperature gradients can turn filament loops (``scroll rings") and delay or even reverse the typical curvature-induced shrinkage \cite{pertsov97}. Similar effects were observed in externally applied electric fields \cite{hauser08}. Moreover, our group showed that in the BZ reaction, filaments can be pinned to inert and impermeable inclusions such as rods, beads, torii, and double torii \cite{jimenez09,dutta11,jimenez12,ke14}. Most notably, local pinning can prevent the collapse of scroll rings, induce stationary twist patterns, create topological frustration, and induce the poorly understood self-wrapping of filaments around the inert heterogeneities \cite{jimenez12}. Recently, Jimenez et al. also demonstrated that scroll rings pinned to two small spheres can be liberated by electric fields in processes involving a re-orientation of the scroll ring and the build-up of filament segments of high curvature near the anchors \cite{jimenez13}.

The response of spiral and scroll waves to external perturbations can be described on the basis of the vortex' response function \cite{keener88,biktasheva10,biktasheva15}. These functions are the eigenfunctions of the adjoint linearized operator corresponding to the critical eigenvalues $\lambda = 0, \pm i \omega$. For most models, response functions are localized to small regions around the spiral core or the filament (see \cite{biktasheva05} for exceptions) which also explains the particle-like behavior of excitation vortices. In 2015, Biktasheva et al. \cite{biktasheva15} used this approach to analyze the drift of scroll waves in thin systems with sharp thickness variations. Such conditions are relevant to certain cardiac arrhythmia because the heart shows complicated thickness variations that potentially affect the dynamics of reentrant waves \cite{yamazaki12}. Their analysis revealed scroll wave drift along steps, ridges, ditches, and disk-shaped thickness variations. For the specific case of an abrupt transition between a medium of height $H_{+}$ and $H_{-}$ (i.e. a step height of $h = H_{+} - H_{-}$), the authors reported a drift speed $dX/dt$ of

\begin{equation}
dX/dt = \epsilon S_{X}(X) , \quad \quad \epsilon = \ln\bigg(\frac{H_{+}}{H_{-}}\bigg),
\end{equation}

\noindent where $S_{X}(X)$ is one spatial component of the response function. As expected, the drift direction of the filament along the edge depends on the chirality of the scroll wave.

In this Article, we present experiments with steps in thin layers of the BZ reaction that verify the theoretical prediction in Eq.~(1). We also observe the step-induced drift of vortices around corners and reproduce the latter finding in numerical simulations. Additional computations show that narrow troughs can induce scroll wave collisions that provoke symmetry breaking repulsion.

\section{Experimental Methods}

Our experimental system consists of a liquid layer of BZ solution in a flat-bottom Petri dish (diameter 5.6~cm). Height variations are created by placing flat glass slides into the Petri dish. No adhesives are used to bond the surfaces because such chemicals could affect the BZ reaction and induce unwanted gradients in excitability along the edge. The viscosity of the reaction medium is increased by addition of xanthan gum (0.4~\% w/v) and agar (0.05~\% w/v) \cite{ke14}. Notice that the system has a free solution-air interface. In all experiments, the initial concentrations of the reactants are: [NaBrO$_3$]~= 62~mmol/L, [H$_2$SO$_4$]~= 175~mmol/L, [malonic acid]~= 48~mmol/L, and [Fe(phen)$_3$SO$_4$]~= 0.0375~mmol/L. The solutions are prepared using nanopure water obtained from a microfiltration system (Barnstead EASYpure UV, 18~$\Omega$cm).

The chemical wave patterns are monitored using a monochrome video camera (COHU 2122) equipped with a dichroic blue filter. The video signal is digitized using a frame grabber (Data Translation DT3155, 640$\times$480 pixels resolution at 8~bit/pixel) and HL Image++97 software. The reaction systems are illuminated by diffuse white light. All measurements are carried out at room temperature. Since our experiments employ a viscous BZ system, scroll waves are readily created by moving a thin glass rod through the solution. This motion breaks spontaneous, non-rotating wave fronts through localized hydrodynamic perturbations. While this procedure allows us to create scroll waves near the step edge, their respective distance varies and does not always induce the desired attraction and drift. The simplicity of the method, however, compensates for this shortcoming as the perturbation can easily be repeated.

\begin{figure}
\includegraphics[width=0.65\linewidth]{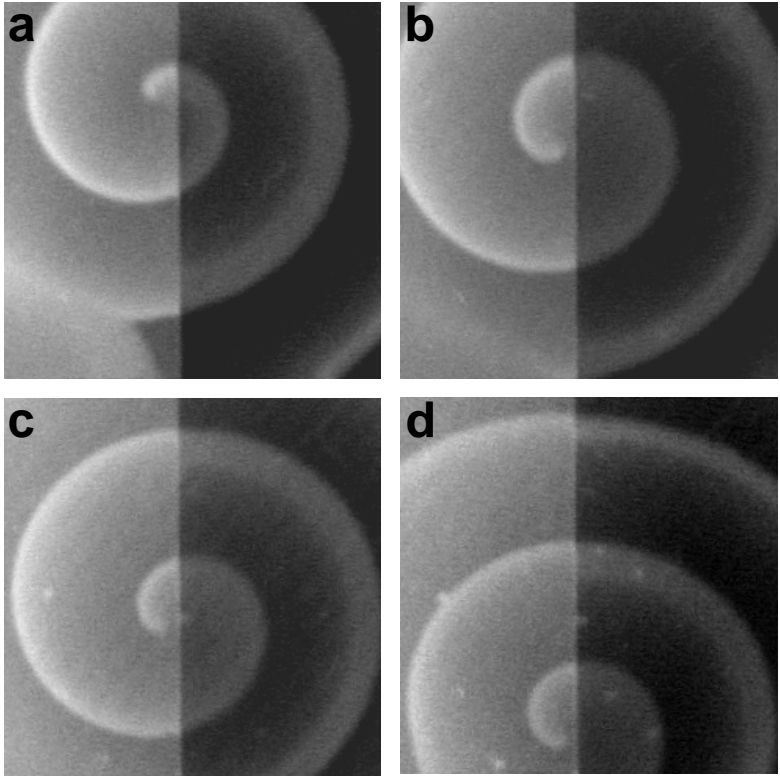}
\caption{\label{fig1} (a-d) Four snapshots of a scroll wave in a three-dimensional layer of BZ solution. The system is thin in the left and thick in the left half. The corresponding step induces drift of the scroll wave center. Time between frames: (a,b) 50~min, (b,c) 50~min, and (c,d) 150~min. Field of view: 1.8~cm~$\times$ 1.8~cm.\\}
\end{figure}

\section{Experimental Results}

All of our experiments are carried out in thin but three-dimensional layers of BZ solution. The system has a positive filament tension and accordingly all filament loops shrink. The thickness of these layers (3.2~mm to 11~mm) is similar to the system-specific wavelength of the unperturbed vortex (4.5~mm). Under such conditions, the filaments are likely to terminate at the upper and lower system boundaries because all other arrangements require a nearly perfect alignment of the filament parallel to these surfaces. Terminating filaments either end at the same or at different surfaces. In the former case, the contracting filament converges to a small half circle that shrinks and annihilates in finite time. In the latter case, the filament converges to a straight line oriented perpendicular to the two boundaries. Furthermore, differences in the rotation phases along the short filament quickly decay \cite{marts08,jimenez09}. Accordingly, a top view of such a thin BZ layer will reveal wave patterns that are essentially identical to quasi-two-dimensional spirals. Such a wave pattern is shown in Fig.~1.

The experimental system in Fig.~1 consists of a thin and a thick layer separated by a sharp step that in the images extends in vertical direction. The height difference between the two regions is created by a glass slide of constant height $h$ (here $h$~= 0.32~cm). The shallow, left side of the image ($H_{-}$~= 0.78~cm) appears on average brighter than the deep, right side ($H_{+}$~= 1.10~cm). This intensity difference is a simple consequence of the Lambert-Beer law that expresses a proportional dependence of light absorbance on optical path length. The dominant light absorbing species in this BZ system is the chemically reduced catalyst ferroin. In propagating wave pulses a significant portion of this compound is oxidized, which reduces light absorbance. Accordingly wave patterns appear as bright regions on a dark background. Absorption changes in vertical direction are not resolved by our experimental set-up. However, this shortcoming compared to tomographic techniques \cite{bansagi06,bansagi07} is easily tolerable for thin layers in which the extent of vertical concentration variations is small.

Figure~1 consists of four consecutive still frames covering a time span of 250~min which is equivalent to about 50 rotation periods. The images show a single, counter-clockwise rotating scroll wave with no or little twist. Its rotation center is located in very close vicinity to the step but clearly remains on the shallow, left side of the system. Most importantly, we observe that the vortex moves along the step line covering a distance of about 1.2~cm which equals more than two wavelengths. This step-induced drift is a verification of the recent theoretical predications by Biktasheva et al. \cite{biktasheva15}. In addition, we observe that the drift commences only if the initial position of the scroll wave is close enough to the step. This is particularly interesting if the vortex center is initially in the deep region. Under this condition, the filament can reach its drift trajectory in the shallow part only if its lower part annihilates at the step wall. We did not further study these initial dynamics because the first two or three rotation cycles of the vortex are typically affected by the methods employed to create the vortex (here hydrodynamic perturbations). Nonetheless, future studies should attempt to analyze the attractor-like features of the step line in more detail.

\begin{figure}
\centering
\includegraphics[width=0.75\linewidth]{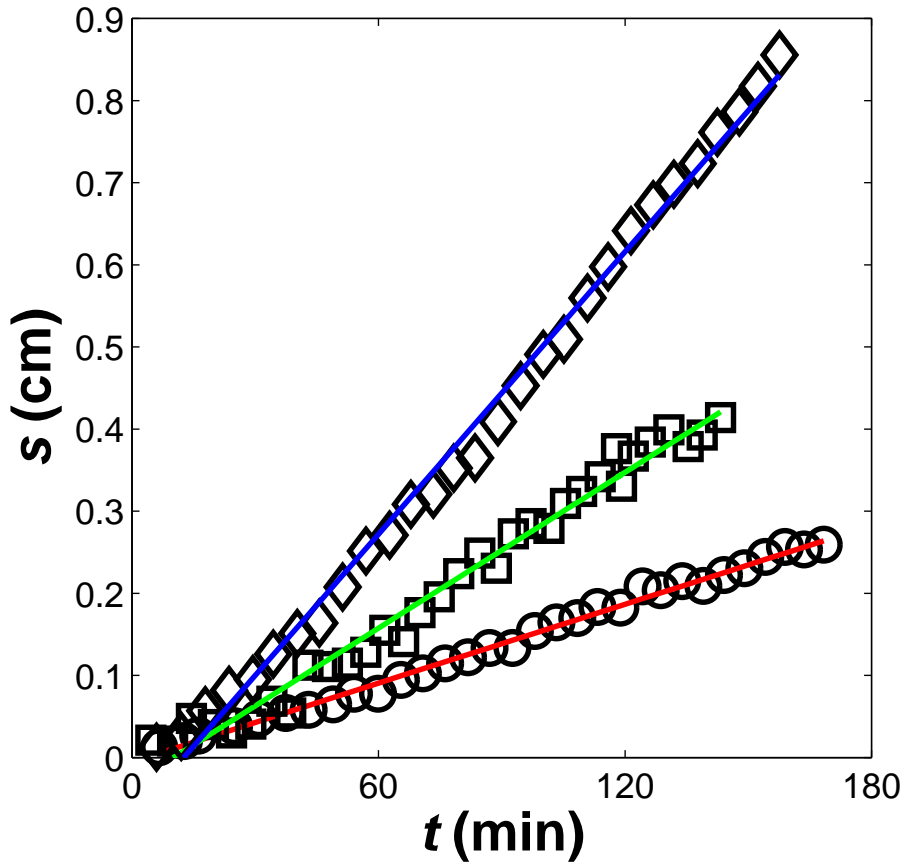}
\caption{\label{fig2} (color online) Experimental data on the temporal evolution of the position of three drifting scroll wave centers. The $s$-coordinate measures space along the linear edge. The step heights are 0.1~cm (diamonds), 0.2~cm (squares), and 0.32~cm (circles). The solution height in the deep half of the system is 1.1~cm. The straight lines are the results are the lines of best fit for the three individual data sets.\\}
\end{figure}

The motion of scroll waves along the line of height change occurs at a constant speed. Figure~2 shows representative measurements of the filament position $s(t)$ for three different step heights and a constant value of $H_{+}$. The three data sets are well described by linear functions and their slopes suggest that the drift speeds increase with increasing step heights. The direction of the drift depends on the chirality of the vortex and on the relative orientation of the shallow and the deep layer. For the arrangement in Fig.~1 (shallow region on the left), we find that clockwise rotating scroll waves move downwards with the respect to the image, whereas counter-clockwise rotating vortices move upwards. We also note that for the BZ system studied here scroll waves exist for more than 9~h. During the late stages of the reaction, chemical processes (foremost the consumption of reactants) induce changes in most if not all system parameters. We specifically find that these changes increase the drift velocity late in the reaction (not shown in Fig.~2).

\begin{figure}
\centering
\includegraphics[width=0.65\linewidth]{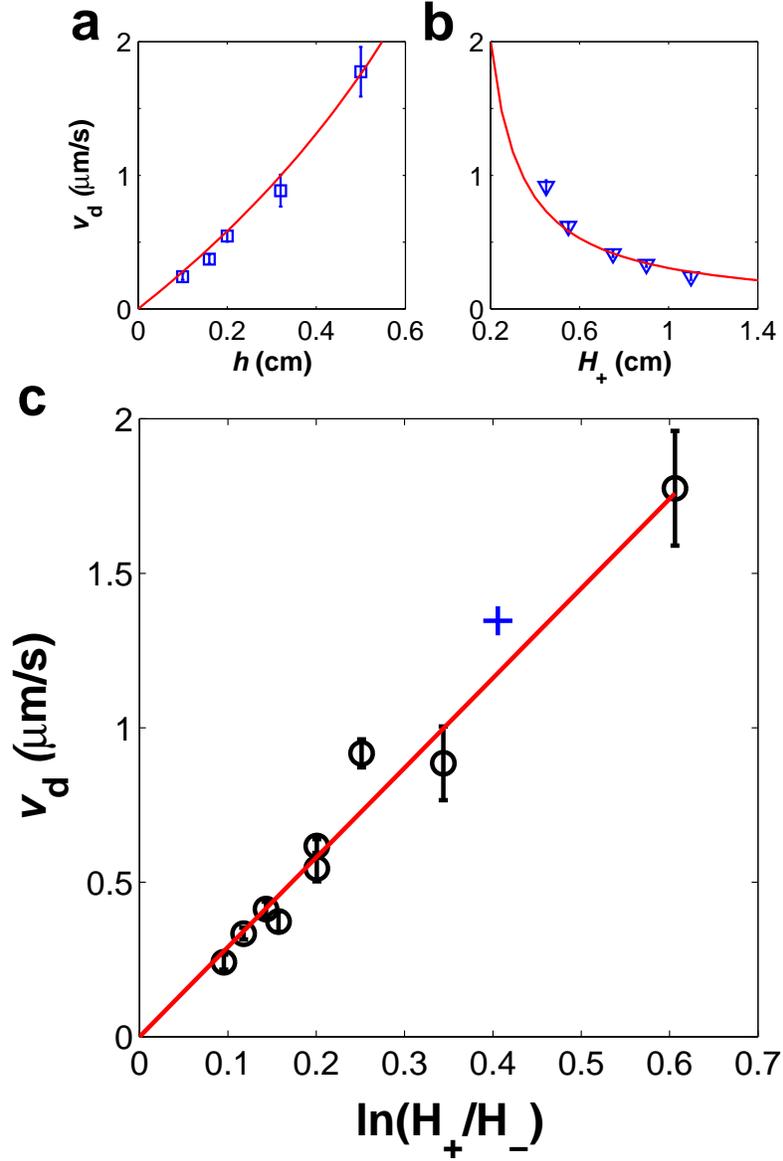}
\caption{\label{fig3} (color online) Drift velocities $v_d$ as a function of the thick ($H_{+}$) and the thin layer height ($H_{-}$). (a) Data obtained at a constant $H_{+}$ of 1.1~cm by variation of the step height $h = H_{+} - H_{-}$. (b) Data obtained for a constant step height $h$ by variation of the solution height $H_{+}$ in the thick system. (c) All drift velocities (circles) are jointly graphed against a logarithmic abscissa. The blue square is obtained from scroll waves drifting along a rectangular trough (see Fig.~4 for details). The red lines are based on the best fit of the proportional dependence $v_d \propto \textrm{ln}(H_{+}/H_{-})$ to the entire data set.\\}
\end{figure}

In the following, we present systematic measurements of the drift velocity $v_d$ for a range of step heights and layer thicknesses. All experiments are performed at least in triplicate and the experimental error bars represent the corresponding standard deviations. Notice that the step height obeys $h = H_{+} - H_{-}$. Figure~3a shows the drift speed as a function of $h$ for $H_{+}$~= 1.10~cm~= \textit{constant}. Over the range of step heights studied, the speed increases with increasing $h$ by a factor of nearly 10. However, an increase in the height of the thick layer at constant step height reveals a monotonic decrease (Fig.~3b). In Fig.~3c, we combine these data and graph the $v_d$ values as a function of $\textrm{ln}(H_{+}/H_{-})$. The experimental results are in excellent agreement with the proportional dependence Eq.~(1) predicted by Biktasheva et al. \cite{biktasheva15}. The proportionality constant is found as $S$~= 2.89~$\mu\textrm{m/s}$. This values allows us to also compute the corresponding graphs for the data in Figs.~3a,b. As expected we again find excellent agreement with Eq.~(1).

\begin{figure}
\centering
\includegraphics[width=0.7\linewidth]{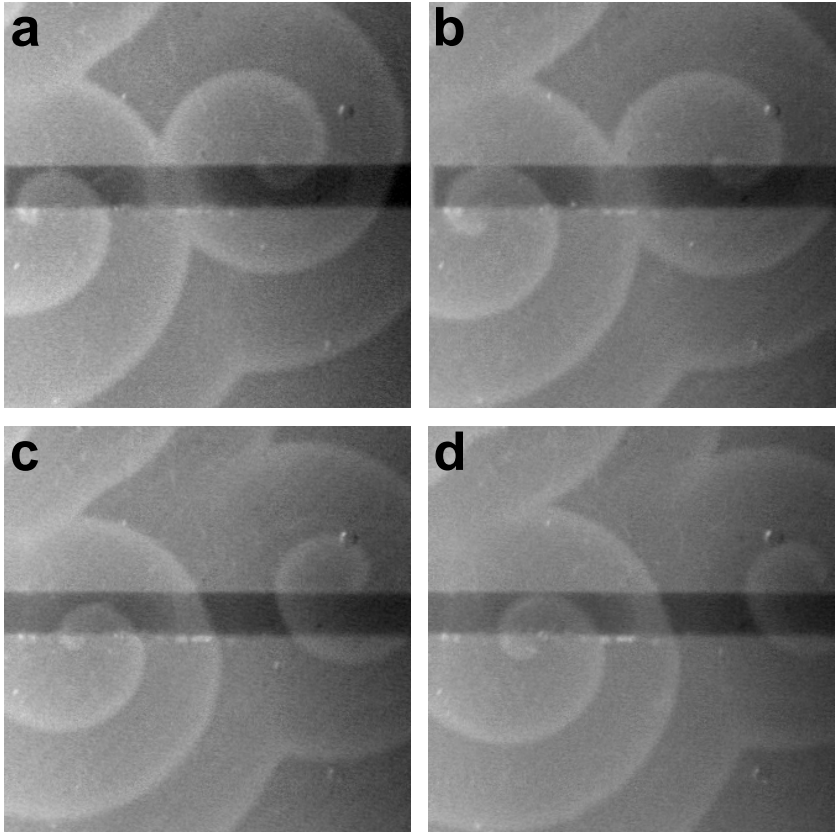}
\caption{\label{fig4} (a-d) Four snapshots of scroll waves drifting along the edge of rectangular trough. Time between consecutive frames: 27~min. Field of view: 2.2~cm~$\times$ 2.2~cm.\newline\newline\newline}
\end{figure}

The height-induced drift of scroll waves is not limited to the case of a linear step but---according to the predictions in \cite{biktasheva15}---also occurs along the border of disk-shaped plateaus and rectangular as well as V-shaped troughs. We have tested one of these more complex cases experimentally. Figure~4 illustrates the dynamics of scroll waves in close vicinity to rectangular trough. The trough is constructed from two glass slides of equal thickness and extends (with respect to the images) in horizontal direction. Its depth below the surrounding plateaus measures 3.2~mm whereas the system thickness in the tough is 9.0~mm. Accordingly, the trough walls have a height of 5.8~mm. The width of the channel equals 2.5~mm which corresponds to approximately 0.5 wavelengths of the unperturbed vortex. Despite this small width, Fig.~4 provides unambiguous evidence for the vortex drift along the trough. In addition, we find that scroll waves of opposite chirality can propagate in the same direction as long as they are affected by different sides of the channel. In Fig.~4, for instance, the clockwise rotating vortex moves rightwards along the upper edge of the trough while the counter-clockwise rotating structure moves also rightwards but in this case along the lower edge. As in the earlier geometries, the filament resides in the shallow regions of the system.

\begin{figure}
\centering
\includegraphics[width=0.7\linewidth]{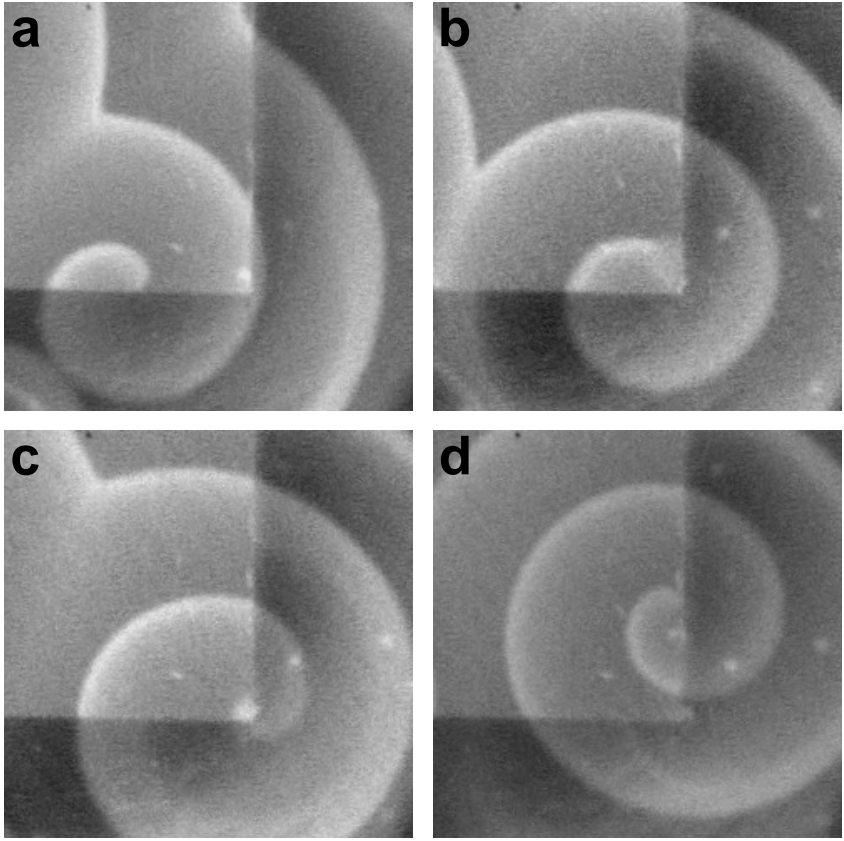}
\caption{\label{fig5} (a-d) Four snapshots of a scroll wave drifting along the edge and around the corner of a rectangular plateau. The system is thin in the upper left region. Time between frames: (a,b) 60~min, (b,c) 8~min, and (c,d) 70~min. System heights: $H_{+}$~= 9.0~mm and $H_{-}$~= 5.8~mm. Field of view: 1.8~cm~$\times$ 1.8~cm.\\}
\end{figure}

Lastly, we report a phenomenon that was not considered in the work by Biktasheva et al. \cite{biktasheva15}. The image sequence in Fig.~5 shows that step-induced vortex drift continues along sharp corners. In this experiment, the step and its corner were created simply by placing a rectangular glass plate of constant height into the reaction medium. We emphasize that these dynamics were observed in numerous experiments and occur reliably. Within the resolution power of our experiment, we find that drift is neither delayed nor accelerated when the scroll wave center approaches and passes the corner. However, we occasionally observe the formation of an unusual excitation wave that emanates from the corner and in some cases rapidly advances the filament along its drift path (see movie in the Supplemental Material). This single excitation pulse might be related to a phase wave triggered by a Doppler-induced local decrease in wave frequency. However, a more likely explanation is the formation of an intermittent, horizontal filament line. This filament would result from the high curvature of the emitted scroll wave front near the forward edge of the corner. Vortex generation near sharp corners has been observed in two-dimensional systems \cite{agladze94} but to our knowledge is an unexplored topic for scroll waves.

\section{Model and Numerical Methods}

Our numerical investigations aim to reproduce the vortex drift around a sharp corner. In addition, we revisit the rectangular trough system and study the possibility of vortex interaction for this geometry. Our simulations are based on the dimensionless Barkley model \cite{barkley02}

\begin{align}
   \addtocounter{equation}{1}
   \frac{\partial u}{\partial t} &= D_u {\nabla}^2 u + {1 \over \epsilon}
       \left\{u(1-u)\left(u-{{v+b}\over{a}}\right)\right\},\tag{2a} \\
\vspace{-0.5cm}
   \frac{\partial v}{\partial t} &= u - v,\tag{2b}
\end{align}

\noindent for which the spatio-temporal dynamics of $u$ and $v$ qualitatively capture the more complicated kinetic behavior of the BZ species bromous acid and the oxidized catalyst (ferriin), respectively. The diffusion coefficient $D_u$ is kept constant at 1.0. The additional model parameters are chosen as $\epsilon$~= 0.025, $a$~= 0.7, and $b$~= 0.01 generating an excitable system with positive filament tension. For the numerical integration of Eqs.~(2), we use forward Euler integration at a time step of $0.006$ and compute the Laplacian using a seven-point stencil. The spatial resolution is 0.2 space units throughout the simulated volume which typically measures 200$\times$200$\times$60 grid points. The system walls, steps, and troughs are no-flux boundaries. Scroll waves are initiated from a layer with high $u$ values and an adjacent refractory layer with high $v$ values. This initial condition creates the filament(s) close to the free edge(s) of the layers. The three-dimensional wave fields are visualized using Matlab.

\begin{figure}
\centering
\includegraphics[width=0.7\linewidth]{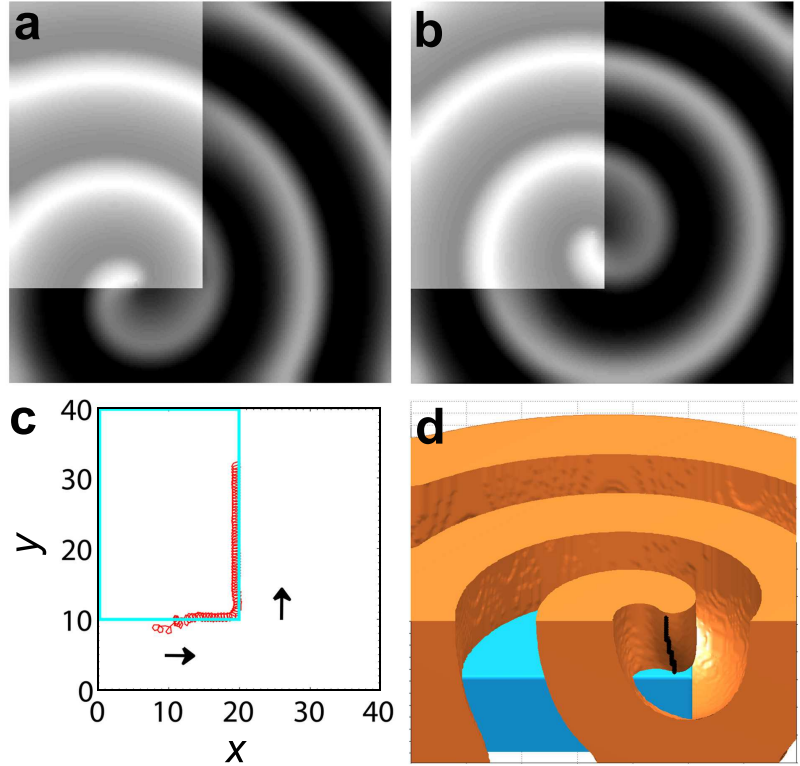}
\caption{\label{fig6} Numerical simulation of a scroll wave drifting around a corner. (a,b) Still images showing the top view of the thin system. The shallow domain is the brighter, rectangular region. Time between frames: 96 time units. (c) Filament trajectory (red) in a plane slightly above the plateau (cyan contour). The arrows indicate the drift direction. (d) Three-dimensional wave field ($u$, orange) and its short filament (black curve) during the turn around the plateau (cyan) corner.}
\end{figure}

\section{Numerical Results}

Figure~6 illustrates a typical simulation in which a scroll wave drifts along the boundary of a plateau-like height variation (shallow region in the upper left portion of (a,b)) and then successfully turns around a corner of that plateau to continue its drift in the upward direction. No unusual extensions or jumps of the filament are observed in any of our simulations. The gray-scale images in Figs.~6a,b are constructed by summation of $v_{x,y}(z)$ over all "vertical" $z$ values. A high value of $v$ can be interpreted qualitatively as a high concentration of the oxidized BZ catalyst which is known to absorb less light than the reduced one. Accordingly $v$ is a rough measure of the transmitted light and we assign the "empty" step region a high and constant transmission. The latter procedure results in the overall brighter gray values in the upper right hand region yielding good agreement with the experimental images (e.g. Fig.~5).

The trajectory of the filament is represented in Fig.~6c. Since the entire filament is oriented nearly perpendicular to the upper and lower system boundaries, the filament motion can be traced by following its cross-section with a plane right above the plateau. The initial filament position (left starting point of red curve) is located in the thicker part of the system. The close vicinity of the step attracts the filament and its lower portion annihilates at the vertical wall of the step. The shortened filament---now connecting the plateau surface and the upper system boundary---traces in very close distance the edge of the step. This distance is slightly increased while turning around the corner. Figure~6d shows a still frame of the three-dimensional wave field during this change in drift direction. Solid (orange) areas correspond to high $u$ values whereas transparent regions indicate low $u$ values. The plateau and the short filament are shown in cyan and black, respectively.

We also performed several simulations for systems subdivided into three rectangular domains with heights $H_i$. Two of these subsystems were aligned to give rise to a T-shaped step arrangement (as viewed from the top). According to Eq.~(1), vortex drift in these system can occur at three different velocities that correspond to the different height ratios along the T-shaped border. Perhaps naively, one can envision a situation in which a scroll wave near the T-junction follows different paths dependent on these specific speeds. However, we found that the filaments always followed the plateau edge to which they initially attached and never spontaneously extended their length to follow a faster or slower edge line.

\begin{figure}
\centering
\includegraphics[width=0.7\linewidth]{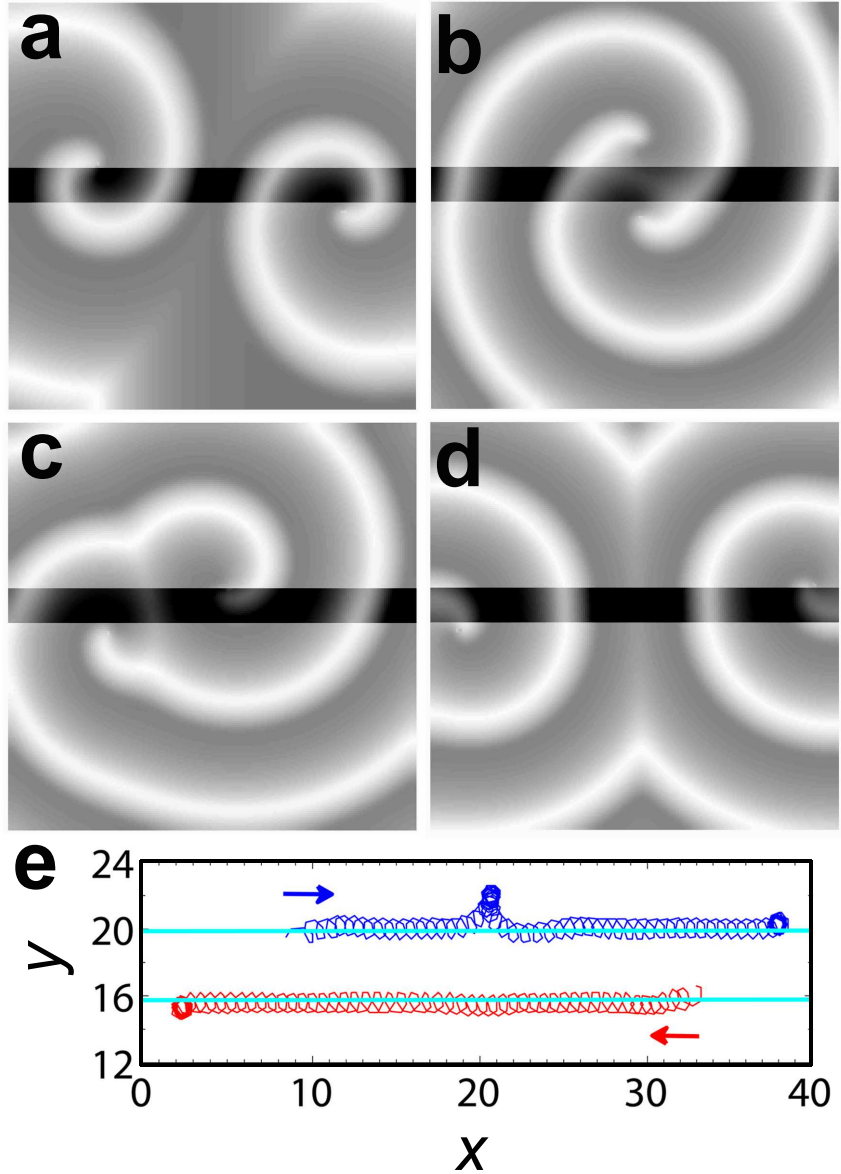}
\caption{\label{fig7} Numerical simulation of two scroll waves drifting along the two edges of a narrow rectangular trough. The simulation reveals repulsive interaction and symmetry breaking. (a-d) Consecutive top views of the wave fields. The trough is the darker stripe. Time between frames in dimensionless time units: (a,b) 73, (b,c) 55, and (c,d) 165. (e) Filament trajectories of the two scroll waves (red and blue lines) monitored in a plane located slightly above the trough (cyan contour). Arrows indicate the respective drift directions.}
\end{figure}

Lastly, we revisited the computational system geometry of a rectangular trough that was original considered by Biktasheva et al. \cite{biktasheva15}. In our simulations, however, we used the scroll wave drift along the edges of the channel to induce vortex interaction. A typical example is shown in Fig.~7. The trough width is significantly smaller than the vortex wavelength and extends (with respect to the image orientation) in the horizontal direction. In (a) two clockwise rotating scroll waves are located at opposite ends of the channel. After a short time, these vortices attach spontaneously to different edges of the trough and begin to drift towards each other (b). As their distance becomes similar to the channel width, the scroll waves interact and, despite a nearly symmetric initial condition, one vortex is pushed away from the trough while its counterpart continues to drift. Once the distance has increased again, the repelled rotor is once again attracted towards the trough edge and also continues its original motion. The corresponding filament trajectories, as measured close to the plateau surface, are shown in Fig.~7e and clearly reveal the existence of a perturbed (upper graph, blue) and an essentially unperturbed path (lower graph, red). We believe that this geometry is a good test ground for systematic analyses of filament interaction.

\section{Conclusions}

For more than three decades, scroll waves have been studied in systems that carefully minimized the influence of the system geometry on the wave dynamics. However, it is quite obvious that in biological systems such as cell clusters or the human ventricles, thickness variations and other geometric features introduce strong perturbations. In this context, we reported here the first experimental examples of scroll wave drift induced by step-shaped height changes that can extend along linear paths or even involve sharp corners. The drift direction is found to depend on the chirality of the vortex and for the case of a narrow trough can translate the vortex center in different directions. In random media, this feature is likely to induce highly dynamic vortex fields that are at least in parts driven by short-range, repulsive vortex interaction and annihilation events.

Perhaps the most important result of our study is the experimental verification of recent predictions by Biktasheva et al. \cite{biktasheva15}. We reemphasize that we obtained quantitative agreement with the predicted proportional dependence of the drift velocity on the logarithm of the ratio of the two system heights. This excellent agreement is not obvious since their analysis is based on asymptotic theory. In addition, we obtained evidence for the existence of an attractive ``force" that induces the collision and shortening of deep-domain filaments at the step (if their distance is sufficiently small). Future experiments could aim to reproduce the predicted drift along the perimeter of small disk-shaped plateaus and explore scroll wave dynamics in systems with complex height variations such as random and periodic patterns.

\section{Acknowledgement}

This material is based upon work supported by the National Science Foundation under Grant No. 1213259. This paper is dedicated to Irving Epstein on the occasion of his 70th birthday.

\end{document}